\documentclass[runningheads]{llncs}
\usepackage{graphicx}
\usepackage{wrapfig}
\usepackage{xcolor}
\usepackage{caption}
\usepackage{subcaption}
\usepackage{microtype}
\usepackage{enumitem}
\usepackage{mathtools}
\usepackage{IEEEtrantools}
\usepackage{amssymb}
\usepackage{pifont}
\usepackage{siunitx}
\usepackage{booktabs,diagbox}
\usepackage{makecell}
\usepackage{multirow}
\usepackage{multicol}
\usepackage{hyperref}

\newcommand{\bftab}{\fontseries{b}\selectfont}

\begin{document}
\title{ELASTIC: Event-Tracking Data Synchronization in Soccer Without Annotated Event Locations}
\titlerunning{ELASTIC}
%
%
\author{
    Hyunsung Kim\inst{1,2}\orcidID{0000-0002-6286-5160} \and
    Hoyoung Choi\inst{1} \and
    Sangwoo Seo\inst{1} \and \\
    Tom Boomstra\inst{3} \and
    Jinsung Yoon\inst{2} \and
    Chanyoung Park\inst{1}
}
\authorrunning{H. Kim et al.}
%
\institute{
    KAIST, Daejeon, South Korea \\
    \email{\{hyunsung.kim,chy3724,tkddn8974,cy.park\}@kaist.ac.kr} \and
    Fitogether Inc., Seoul, South Korea \\
    \email{\{hyunsung.kim,jinsung.yoon\}@fitogether.com} \and
    AFC Ajax, Amsterdam, Netherlands \\
    \email{t.boomstra@ajax.nl}
}
\maketitle              
\begin{abstract}
The integration of event and tracking data has become essential for advanced analysis in soccer. However, synchronizing these two modalities remains a significant challenge due to temporal and spatial inaccuracies in manually recorded event timestamps. Existing synchronizers typically rely on annotated event locations, which themselves are prone to spatial errors and thus can distort synchronization results. To address this issue, we propose ELASTIC (\textbf{E}vent-\textbf{L}ocation-\textbf{A}gno\textbf{STIC} synchronizer), a synchronization framework that only uses features derived from tracking data. ELASTIC also explicitly detects the end times of pass-like events and separates the detection of major and minor events, which improves the completeness of the synchronized output and reduces error cascade across events. We annotated the ground truth timestamps of 2,134 events from three Eredivisie matches to measure the synchronization accuracy, and the experimental results demonstrate that ELASTIC outperforms existing synchronizers by a large margin.
\end{abstract}

\section{Introduction}
The increasing availability of detailed in-game data has revolutionized the way of analyzing matches and players in soccer. Among the various data modalities, two primary pillars stand out: \textit{event data} and \textit{tracking data}. Event data consists of manually annotated records of on-the-ball events such as passes, shots, and tackles. This form of data has been available relatively earlier and has served as the foundation for a wide range of analytical tasks, including performance evaluation~\cite{BransenV20,DecroosBVD19,LiuLSK20,MerhejBMR21,PappalardoCFMPG19,Singh19} and tactical analysis~\cite{BealCNR20Optimising,ClijmansRD22,DecroosHD18,VanRoyRYRD23}. On the other hand, tracking data provides spatiotemporal information on all players and the ball at every moment of the game, collected via optical or wearable sensor-based systems. With recent advances in tracking technologies, such data have become increasingly widespread, allowing for richer analyses that consider the movement context of multiple players surrounding each event~\cite{AnzerB21,AnzerB22,FernandezB20,FernandezBC19,FernandezBC21,RahimianKST23,RahimianVAT22,RahimianVT23,RobberechtsVD23,Spearman18,SpearmanAGRP17}.

However, aligning these two modalities is nontrivial. In practice, manually recorded event timestamps often suffer from temporal inaccuracies. When naively combined with tracking data, such imprecision can distort the contextual understanding at the time of events, leading to unreliable downstream analyses.

\begin{figure}[!b]
\centering
\subfloat[Snapshot\label{fig:sample_snapshot}]{
	\includegraphics[width=0.42\textwidth]{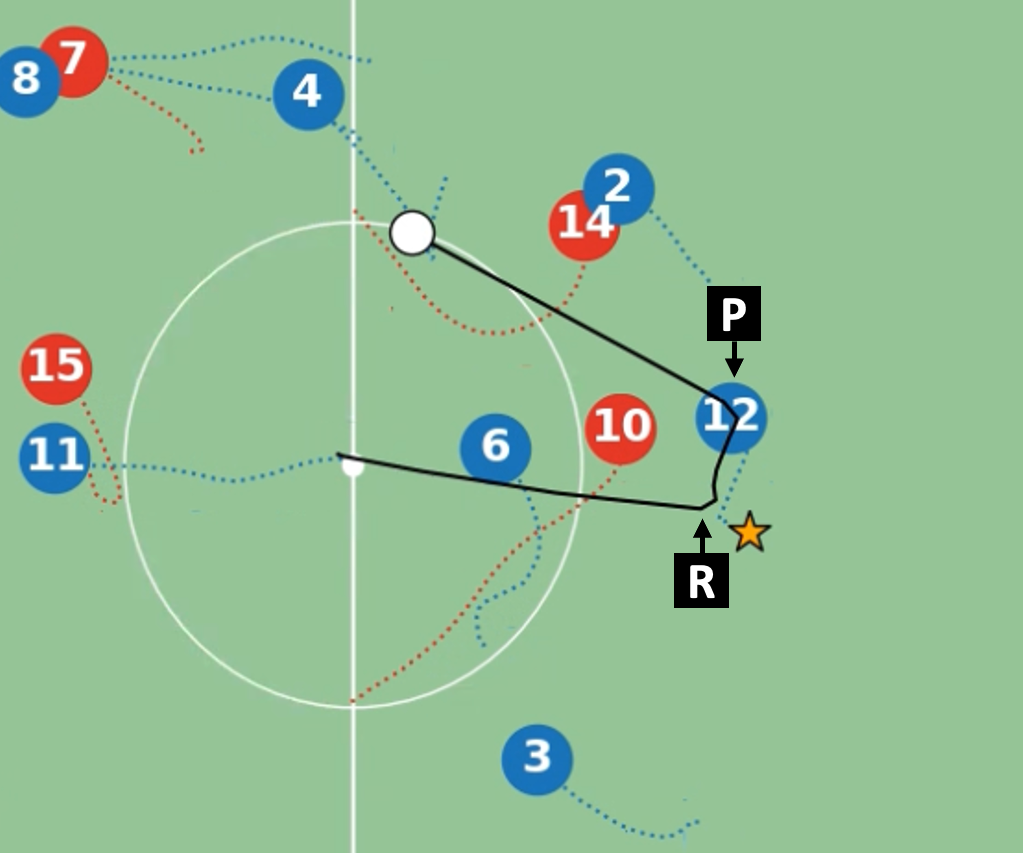}
}
\subfloat[Distance values from the ball\label{fig:sample_plot}]{
	\includegraphics[width=0.48\textwidth]{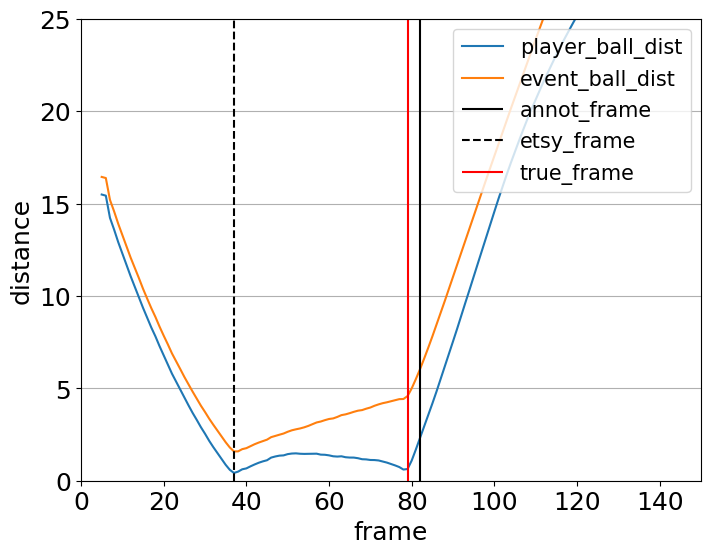}
}
\caption{The snapshot and features of a pass by player \#12. The white circle with a black tail (left) shows the ball trajectory, and the star marks the annotated pass location. The blue and orange plots (right) indicate the distance of the player and the annotated event from the ball, respectively. The black and red solid lines are annotated and true pass frames, and the black dashed line is the frame matched by existing methods~\cite{AnzerB21,VanRoyCD23}. Since the annotated pass location (star) is closer to the earlier receive point (R) than the true pass location (P), the methods wrongly select the earlier receive frame as the estimated pass timestamp.}
\label{fig:sample}
\end{figure}

To address this, several methods for synchronizing soccer event and tracking data have been proposed. Biermann et al.~\cite{BiermannKRMEM23} introduced a learning-based approach that applies a sliding window over tracking data and uses aggregated window features to classify whether each window's center frame corresponds to a pass. However, it requires labeled training data to obtain prototype features for pass and non-pass windows, and its reliance on coarse window-level features rather than frame-level temporal cues limits its accuracy (e.g., fewer than 10\% of passes were aligned exactly to the correct frame). Other studies~\cite{AnzerB21,VanRoyCD23} proposed distance-based methods that define a window around the annotated event timestamp and find the frame that minimizes the sum of distances between the annotated event location, ball location, and event player location. However, their reliance on the annotated event locations, which are also prone to spatial error, can lead to frequent misalignments as illustrated in Fig.~\ref{fig:sample}. Moreover, these errors can propagate across subsequent events since they attempts to preserve event order. These issues have been largely overlooked in the prior studies due to the absence of thorough evaluation with ground truth.

Another neglected challenge is the lack of explicit ball-receiving events in most event datasets. While the datasets often include annotated end locations of events, existing methods focus only on aligning the events' initiation moments. Consequently, their end times remain undetected, and the annotated end locations also remain uncorrected. This hinders the accurate reconstruction of ball drives and degrades the reliability of models that rely on precise pass destinations, such as Pitch Control~\cite{Spearman18,SpearmanAGRP17}, EPV~\cite{FernandezBC19,FernandezBC21}, xPass~\cite{AnzerB22}, or un-xPass~\cite{RobberechtsVD23}.

To overcome these challenges, we propose a novel synchronization framework, ELASTIC (\textbf{E}vent-\textbf{L}ocation-\textbf{A}gno\textbf{STIC} synchronizer), which infers true event timings using only tracking-based features. Its key innovations are threefold:
\begin{enumerate}
    \item It leverages features such as distances between players and the ball, pre- and post-event kick distances, and ball acceleration to determine the most plausible event moment within a temporal window, without relying on noisy annotated event locations.
    \item It explicitly detects the end times of pass-like events, enabling precise determination of ball drives and supporting a wider range of analytical tasks.
    \item It separates the detection of major (such as passes, shots, interceptions, and set-pieces) and minor events (such as take-ons, tackles, fouls, and bad touches) to mitigate error cascade. Specifically, ELASTIC first detects major events and their end times. Then, minor events are searched only within the interval between two adjacent events. This staged approach ensures that even if minor events are inaccurately detected, their errors do not cascade to the detection of subsequent events.
\end{enumerate}

To facilitate rigorous evaluation, we manually annotated ground truth timestamps of 2,134 events from the first halves of three Dutch Eredivisie 2024–25 matches. Experimental results show that ELASTIC significantly outperforms existing synchronization methods. In particular, while the previous state-of-the-art method, ETSY~\cite{VanRoyCD23}, achieved only 21.1\% exact alignment, ELASTIC dramatically improved this to 88.4\%, highlighting its strong effectiveness. The source code is available at \url{https://github.com/hyunsungkim-ds/elastic.git}, along with a tutorial for applying ELASTIC to Sportec Open Dataset\footnote{\url{https://springernature.figshare.com/articles/dataset/An_integrated_dataset_of_spatiotemporal_and_event_data_in_elite_soccer/28196177}}~\cite{BassekRWM25} that contains event and tracking data from seven matches of German Bundesliga's first two divisions.

\section{Proposed Framework} \label{se:framework}
Given a pair of event and tracking data from a soccer match, our objective is to infer the true start (and end) timestamps of each on-the-ball event. The event data provides each event's annotated timestamp, the player involved, the event type, and its outcome. Notably, unlike prior studies~\cite{AnzerB21,VanRoyCD23}, our framework does not require the annotated event location and its body part. The tracking data contains spatiotemporal information of 10 or 25 frames per seconds (FPS), where each frame data consists of $(x,y)$ coordinates of every player, $(x,y,z)$ coordinates of the ball, and the flag indicating whether the frame belongs to an open play or a dead ball phase. With this pair of data, we aim to assign each event to a specific frame during open play. For pass-like events, we additionally want to detect their end frame, corresponding to the moment the ball is received or goes out of play. The definition of pass-like events is discussed in Section \ref{se:preproc}.

\begin{figure}[t!]
    \centering
    \includegraphics[width=0.6\textwidth]{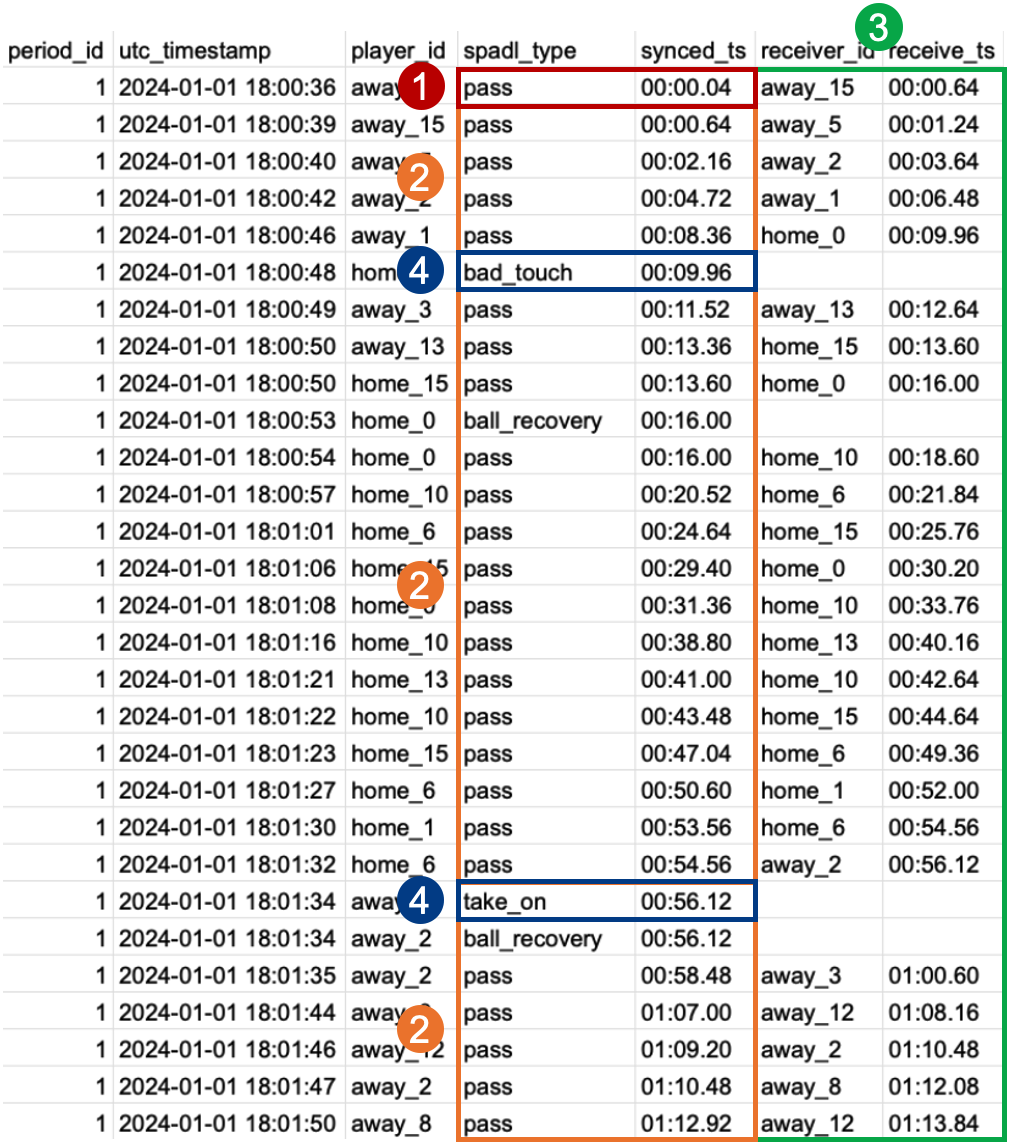}
    \caption{Synchronization stages of ELASTIC.}
    \label{fig:stages}
\end{figure}

As instantiated in Fig.~\ref{fig:stages}, the proposed ELASTIC consists of the following four stages: data preprocessing and kick-off synchronization (Section \ref{se:preproc}), major event synchronization (Section \ref{se:major_events}), receive detection for pass-like events (Section \ref{se:receive}), and minor event synchronization (Section \ref{se:minor_events}). The subsequent sections detail the algorithmic design and rationale for these stages.

\subsection{Data Preprocessing and Kick-off Synchronization} \label{se:preproc}
First, we preprocess the input event and tracking data and remove the constant time offset between them. Unless otherwise noted, we follow the preprocessing and kick-off detection procedures of ETSY~\cite{VanRoyCD23}, including transformation into the $[0,105] \times [0, 68]$ IFAB coordinate system, aligning the direction of play so that the home team attacks from left to right, removing dead ball phases, and calculating velocity and acceleration values of players and the ball. In particular, we adopt the SPADL~\cite{DecroosBVD19} representation to standardize event types and the \texttt{kloppy}~\footnote{\url{https://kloppy.pysport.org}} format for tracking data. After data preprocessing, kick-off frame is determined for each of the period using the player and ball locations and ball acceleration.

However, we introduce three key modifications in data preprocessing to extend coverage and improve the quality of motion features. First, while we follow SPADL definitions for event categorization, we additionally include the following four types to support broader coverage of in-game actions: \textit{ball recovery} when a player gains possession of an uncontrolled ball, \textit{dispossessed} when a player loses possession while dribbling, \textit{shot block} when a defender blocks an opponent's shot, and \textit{keeper sweep} when the goalkeeper leaves the penalty area to clear the ball. With these additions, the event types are grouped into the following categories:
\begin{itemize} 
    \item \textbf{Open-play pass-like:} pass, cross, shot, clearance, keeper punch, \textit{shot block}
    \item \textbf{Set-piece pass-like:} throw-in, goal kick, short/crossed corner kick, short/crossed free kick, free kick shot, penalty shot
    \item \textbf{Incoming:} interception, keeper save/claim/pick-up/\textit{sweep}, \textit{ball recovery}
    \item \textbf{Minor:} tackle, foul, bad touch, take-on, \textit{dispossessed}
\end{itemize}
Among these categories, all except minor are synchronized in Section \ref{se:major_events} and pass-like events in open play and set-piece are further extended to detect their end frames in Section \ref{se:receive}. Lastly, minor events are synchronized in Section \ref{se:minor_events}.

The second difference lies in how acceleration is defined and processed. Rather than using the change in scalar speed (termed \textit{speed-based acceleration} hereafter) as in ETSY, we compute \textit{acceleration} as the norm of the velocity vector's derivative to capture changes not only in speed but also in movement direction. While ETSY considers only the sign of speed-based acceleration to filter out frames, we restrict candidate frames to only the peaks in \textit{velocity-based} acceleration, thereby narrowing the search to true moments of abrupt directional change in ball movement. However, this approach is sensitive to positional noise in tracking data, which can amplify acceleration values and produce numerous false peaks. To address this, we apply a Savitzky–Golay filter~\cite{SavitzkyG64} to both velocity and acceleration profiles, as in previous studies for modeling sports trajectories~\cite{ChoiKLJKYK25,KimCKYK23}.

The last difference is the explicit \textit{episode} segmentation to improve the detection accuracy of set-piece events. Most set-pieces excluding fast restarts in some throw-ins are expected to resume play following a dead ball phases. To utilize this fact in our synchronizer in Section \ref{se:major_events}, we follow Kim et al.~\cite{KimCKYK23} by defining an \textit{episode} as a sequence of consecutive open-play frames without interruption and segmenting the match into the resulting episodes.

\subsection{Major Event Synchronization} \label{se:major_events}
After aligning the kick-off frame in each half, we proceed to synchronize the timestamps of pass-like (including both open play and set-piece) and incoming events in chronological order. Although our feature engineering and scoring strategies differ significantly from those of ETSY~\cite{VanRoyCD23}, the overall process follows a similar three-step process: (1) defining a \textit{qualifying window}, (2) extracting \textit{candidate frames}, and (3) selecting the \textit{best frame} based on scoring functions.

\paragraph{Qualifying window detection.} For each event, we first define a temporal window within which the true timestamp likely falls. For set-piece event types except throw-in, the window spans from the start of the corresponding episode to one second later, based on the assumption that set-pieces typically begins new episodes. For all other events, we define the qualifying window as a symmetric 10-second window centered on the annotated event timestamp.

\paragraph{Candidate frame extraction.}
Inside the qualifying window, we filter frames to select only those that satisfy the following criteria:
\begin{itemize}
    \item The frame is after the synchronized timestamp of the previous event.
    \item The player-ball distance, i.e., the distance between the player executing the event and the ball, is less than \SI{3}{m} and the ball height is less than \SI{3.5}{m}.
    \item The frame should be either (1) the local minimum of the player-ball distance signal, (2) the local minimum of the ball height signal, or (3) the local maximum of the smoothed ball acceleration signal.
\end{itemize}

\paragraph{Best frame selection.}
For candidate frames $\{t_1, \ldots, t_N\}$ resulting from the previous step and the annotated frame $t_a$ of a given event, we score each candidate $t_k$ based on the following temporal features derived from tracking data. All scoring functions use a clipped linear mapping defined as
\begin{equation}
    f(x;x_0,x_1) =
    \begin{cases}
        0 & \text{if } x \le x_0, \\
        1 & \text{if } x \ge x_1, \\
        \frac{x - x_0}{x_1 - x_0} & \text{if } x_0 < x < x_1.
    \end{cases}
\end{equation}
The detailed description and scoring criterion for each feature are as follows:
\begin{itemize}
    \item \textbf{Player-ball distance $d(t_k)$ (PBD):} Since the executing player touches the ball at the actual moment of event, a large $d(t_k)$ should decrease the score. Thus, we define a decreasing scoring function $s_1$ for the PBD as
    \begin{equation}
        s_1(t_k) = \lambda_1 \cdot (1 - f(d(t_k); 0, 3\text{\,m})).
    \end{equation}
    \item \textbf{Ball acceleration $a(t_k)$ (BA):} Since an event usually triggers a sudden change in ball trajectory, a large $a(t_k)$ should increase the score. Thus, we partially score $t_k$ according to the BA using an increasing function
    \begin{equation}
        s_2(t_k) = \lambda_2 \cdot f(a(t_k); 0, 20\text{\,m/s}^2).
    \end{equation}
    \item \textbf{Post-event kick distance $\max_{t_k < t < t_{k+1}} d(t)$ (Post-KD, for pass-like):} The maximum PBD in the interval between the current candidate $t_k$ and the next candidate $t_{k+1}$. Since a longer Post-KD implies a pass rather than a dribble, it should lead to a high score when detecting a pass-like event. Thus, the scoring function $s_3$ for the Post-KD is defined as
    \begin{equation}
        s_3(t_k) = \lambda_3 \cdot f\left( \max_{t_k < t < t_{k+1}} d(t); 0, 5\text{\,m} \right).
    \end{equation}
    \item \textbf{Pre-event kick distance $\max_{t_{k-1} < t < t_k} d(t)$ (Pre-KD, for incoming):} The maximum PBD in the interval between the previous candidate $t_{k-1}$ and the current candidate $t_k$. Since a longer Pre-KD indicates an actual receive rather than subsequent minor touches, it deserves a high score when detecting an incoming event. Thus, the scoring function $s_4$ for the Pre-KD is defined as
    \begin{equation}
        s_4(t_k) = \lambda_4 \cdot f\left( \max_{t_{k-1} < t < t_k} d(t); 0, 5\text{\,m} \right).
    \end{equation}
    \item \textbf{Frame delay $\max\{t_k-t_a,0\}$ (FD):} In cases where the same player performs multiple events in a qualifying window, earlier candidates are preferred (because we are detecting events in chronological order). Thus, we penalize the candidates at later frames than the original annotated frame using a decreasing scoring function $s_5$ defined as
    \begin{equation}
        s_5(t_k) = \lambda_5 \cdot (1 - f(\max\{t_k-t_a,0\}; 0, 5\text{\,s})).
    \end{equation}
\end{itemize}
In the above definitions, $\lambda_1, \ldots, \lambda_5$ are non-negative scaling coefficients that add up to 100. This leads to the total score $s(t_k) = \sum_{i=1}^5 s_i(t_k)$ falls between 0 and 100 according to its feature values. Since Post-KD and Pre-KD are important only for pass-like and incoming events, respectively, we set the coefficients as $(20, 20, 20, 0, 40)$ for the former and $(20, 20, 0, 20, 40)$ for the latter. Finally, the best frame $t^*$ that maximizes the score among the candidates, i.e.,
\begin{equation}
    t^* = \arg\max \{ s(t_k) \}_{k=1}^N.
\end{equation}

\begin{figure}[!bt]
\centering
\subfloat[Pass\label{fig:feat_pass}]{
	\includegraphics[width=0.45\textwidth]{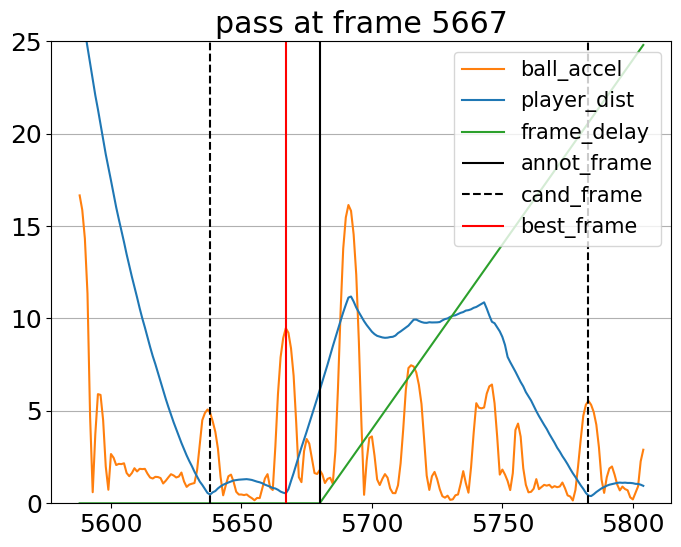}
}
\subfloat[Ball recovery\label{fig:feat_recovery}]{
	\includegraphics[width=0.45\textwidth]{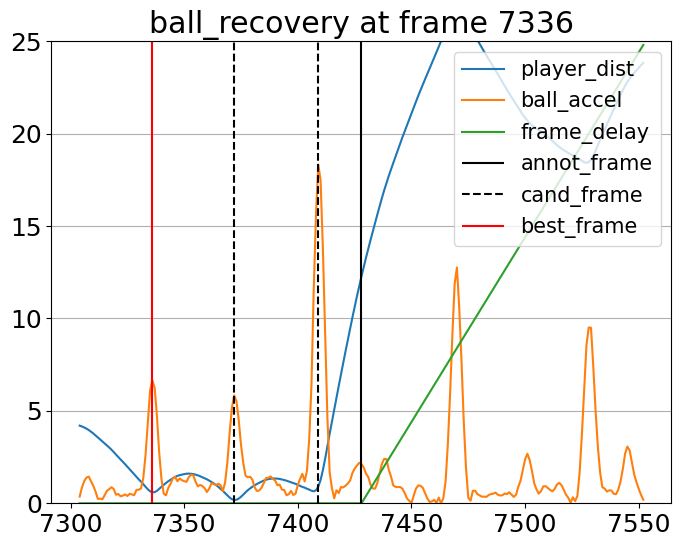}
}\vspace{0.5em}
\subfloat[Receiving a pass\label{fig:feat_receive}]{
	\includegraphics[width=0.45\textwidth]{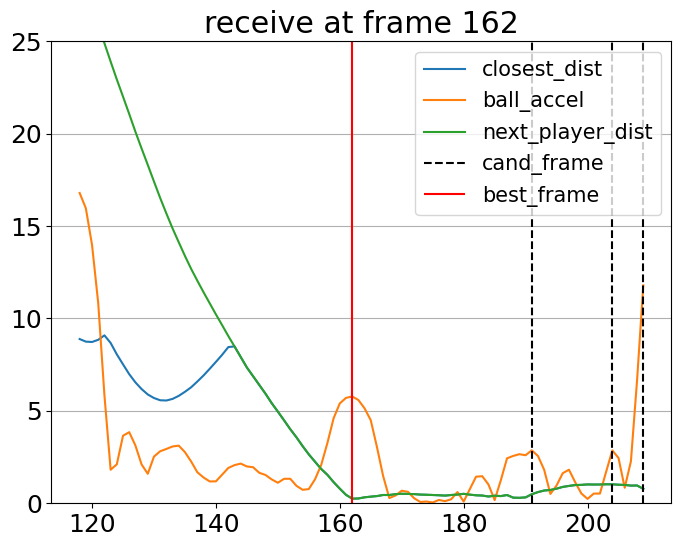}
}
\subfloat[Tackle\label{fig:feat_tackle}]{
	\includegraphics[width=0.45\textwidth]{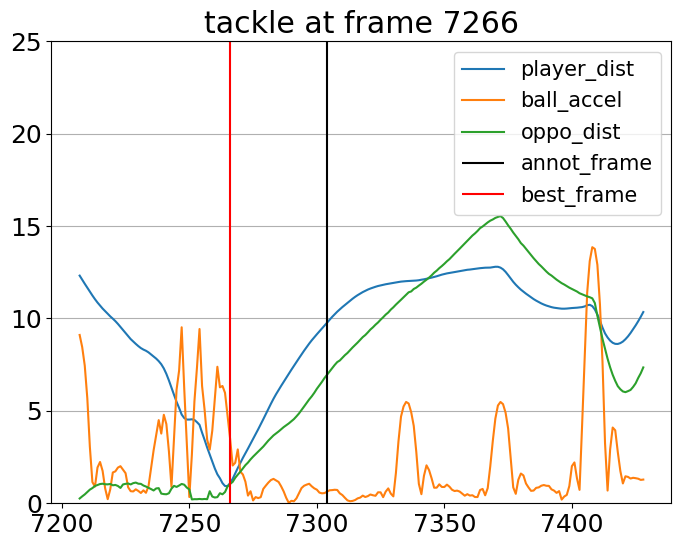}
}
\caption{Feature plots obtained from 25 FPS tracking data around events. BA and FD values are divided by 5 to match the scale of distance values.}
\label{fig:feat_major}
\end{figure}

Fig.~\ref{fig:feat_pass} and \ref{fig:feat_recovery} illustrate the feature values within the qualifying window for a pass-like and an incoming event. In Fig.~\ref{fig:feat_pass}, the first candidate frame (5638) corresponds to the moment of receiving the ball rather than a pass. This frame yields a low score due to a small Post-KD, thereby correctly excluded. In addition, the player makes another touch at frame 5783, but this later action is assigned a high FD value, which penalizes and prevents it from being wrongly detected. In Fig.~\ref{fig:feat_recovery}, many candidate frames (7336, 7372, and 7409) appear as the player dribbles. However, frame 7336, which is the true moment of ball recovery, is correctly identified because of a large Pre-KD. These examples show that each feature contributes meaningfully to disambiguating the true event frame.

\subsection{Receive Detection} \label{se:receive}
For all previously synchronized \textit{pass-like} events (in open play or set-piece), ELASTIC additionally detects their \textit{end frames} and \textit{receivers}. At the end of a pass-like event, one of three situations typically occurs: (1) the ball leaves the field of play, (2) a goal is scored (in the case of a shot), or (3) a player receives the ball. Accordingly, the receiver is labeled as \textit{out}, \textit{goal}, or a player.

The first two cases are straightforward. Following Vidal-Codina et al.~\cite{VidalCodinaEEB22}, if the next event is a set-piece, we label the receiver as \textit{out}; if the current event is a shot and the next event is a pass by the opposing team starting from the center circle, the receiver is labeled as \textit{goal}. In both cases, we assign the last frame of the current episode as the end of the event.

When the ball remains in play, the detection process varies according to the type of the next event. If it is a shot block, keeper punch, or an incoming event that inherently implies a receive, then we directly assign its timestamp as the end frame and its executing player as the receiver. Otherwise, we search for the end frame and receiver jointly within a qualifying window between the current and next event frames.

The search process for the latter largely mirrors incoming event synchronization in Section~\ref{se:major_events} with one key difference: the receiver is unknown, so PBD cannot be directly computed. Instead, based on the distance $d_p(t)$ between the ball and each candidate receiver $p \in P$, we calculate two features at each frame $t$: the \textit{closest player-ball distance} (CPBD) $\min_{p \in P} d_p(t)$ and the \textit{next player-ball distance} (NPBD) $d_{p'}(t)$ where $p'$ is the next event player and $P$ is the set of candidate receivers (i.e., the passer's teammates for successful pass-like events and opponents for failed ones). We use the CPBD instead of PBD when extracting candidate frames, and then compute CPBD and NPBD scores as
\begin{equation}
    s_6(t) = \lambda_6 \cdot (1 - f(\min_{p \in P} d_p(t); 0, 3\text{\,m})), \quad s_7(t) = \lambda_7 \cdot (1 - f(d_{p'}(t); 0, 3\text{\,m})).
\end{equation}
Each candidate frame $t_k$ is scored by $s(t_k) = s_2(t_k) + s_4(t_k) + s_6(t_k) + s_7(t_k)$ where $s_2$ and $s_4$ are the BA and Pre-KD scores introduced in Section~\ref{se:major_events}). Lastly, the candidate with the highest score is selected as the end frame, and the closest player to the ball at that frame is labeled as the receiver. All $\lambda$ values are set to 25. See Fig.~\ref{fig:feat_receive} as an example.

\subsection{Minor Event Synchronization} \label{se:minor_events}
Unlike pass-like or incoming events, the minor events listed in Section \ref{se:preproc} often lack clear observable cues such as sudden ball acceleration or increased player-ball distance, making them more error-prone. To prevent such errors from propagating to the detection of subsequent events, we isolate minor event synchronization into a separate final stage, after all other events have been synchronized.

Each minor event is detected within a window bounded by the end (or start, if undefined) of the previous event and the start of the next synchronized event. We apply type-specific detection rules as follows:
\begin{itemize}
    \item \textbf{Tackle:} As that a tackle typically involves a challenge between the tackler and the ball-possessing opponent, we first identify the \textit{target opponent} as the receiver (if pass-like) or executer (if incoming) of the previous event. Then, we find the best frame by scoring candidate frames based on (1) PBD, (2) BA, and (3) \textit{target opponent-ball distance} (TOBD). As shown in Fig.~\ref{fig:feat_tackle}, the score effectively pinpoints the tackling frame at which the two competing players and the ball become close to each other in a moment.
    \item \textbf{Foul:} Since it triggers a dead ball phase, we assign the end of the current episode as its aligned frame.
    \item \textbf{Bad touch:} If the executing player is the same as the previous event's receiver, we assign the end frame of the previous event. Otherwise, we detect a frame using the same method as the pass-like event synchronization.
    \item \textbf{Take-on:} We define a take-on moment as the point where a player begins to accelerate in an attempt to dribble past a nearby opponent. To detect such moments, we first identify the \textit{target opponent at each frame} as the closest opponent among those who are positioned closer to the target goal than the ball carrier. Since take-ons typically involve a burst of speed and a noticeable shift in the spatial relationship between the player and the defender, we score each candidate frame based on five features: (1) BA, (2) TOBD, (3) \textit{player max speed} (PMS) within the interval up to the next candidate, (4) \textit{player delta speed} (PDS), defined as the difference between PMS and the current speed, and (5) \textit{player-opponent angle change} (POAC), measuring the directional shift in their relative positions before and after the candidate.
    \item \textbf{Dispossessed:} These events often co-occur with tackles. Thus, if the next event is a tackle, we first synchronize the tackle and then align the dispossessed event to the same frame. In cases where dispossessed events occur in isolation, we detect a best frame by scoring candidate frames based on PBD, Post-KD, and \textit{relative ball acceleration} (RBA). The last feature is defined as the change in relative speed between the player and the ball to capture the subtle loss of control or deflection.
\end{itemize}
In particular, ETSY~\cite{VanRoyCD23} does not handle dribbles (including take-on and dispossessed events) at all, despite their importance in many soccer analytics tasks. Moreover, its detection accuracy for tackle, foul, and bad touch is low, and errors in these detections negatively affect the synchronization of other events. In contrast, our method robustly detects minor events by applying tailored, type-specific rules and isolating the detection into a separate stage.

\section{Experiments} \label{se:experiments}
For rigorous evaluation, we first re-annotated the event timestamps in the dataset to obtain accurate ground truth labels and validated them via a cross-annotator comparison (Section \ref{se:annot}). We then evaluated the accuracy of different synchronizers by comparing their outputs to this refined labels (Section~\ref{se:baseline}). Lastly, we conducted an ablation study about our multi-stage strategy that splits major and minor events into distinct synchronization stages (Section~\ref{se:ablation}).

\subsection{Ground Truth Annotation} \label{se:annot}
To obtain ground truth, three authors with expertise in soccer analytics re-labeled 2,134 events from the first half of three matches in Dutch Eredivisie 2024-25. For pass-like and set-piece events, end timestamps were also annotated. All labels were annotated at 25 FPS (\SI{0.04}{s} resolution) to align with the tracking data.

As a result, Table~\ref{tab:annot} shows that over 90\% of event times were labeled with exact agreement across all annotators with the mean difference (MD) under one frame (\SI{0.04}{s}). This high consistency stems from a meticulous re-annotation process, where each annotator paused the video at each event to pinpoint its exact frame. Notably, for nearly 99\% of events had at least two annotations within two frames (\SI{0.08}{s}) of each other. Assuming that such close agreement reflects its reliability, this result implies that the median of the three timestamps for each event provides a reliable estimate with a confidence of 99\%. With this statistical insight, we adopt the median annotated timestamps for each event as the ground truth in subsequent experiments.

\begin{table}[b!]
    \centering
    \resizebox{0.8\textwidth}{!}{
    \begin{tabular}{c|rr|rrrr}
        \toprule
        \textbf{Category} & \makecell[c]{\bf Total} & \makecell[c]{\bf MD} & \makecell[c]{\bf Exact3} & \makecell[c]{\bf Exact2} & \makecell[c]{\bf Close3} & \makecell[c]{\bf Close2} \\
        \midrule
        Pass-like & 1,590 & 0.245 & 1,516 (95.3\%) & 1,582 (99.5\%) & 1,564 (98.4\%) & 1,588 (99.9\%) \\
        Set-piece &   117 & 0.121 &   108 (92.3\%) &   115 (98.3\%) &   113 (96.6\%) &   117 (100.0\%) \\
        Incoming  &   168 & 2.258 &   148 (88.1\%) &   161 (95.8\%) &   154 (91.7\%) &   165 (98.2\%) \\
        Minor     &   259 & 4.900 &   158 (61.0\%) &   216 (83.4\%) &   186 (71.8\%) &   242 (93.4\%) \\
        \midrule
        Event start & 2,134 & 0.961 & 1,930 (90.4\%) & 2,074 (97.2\%) & 2,017 (94.5\%) & 2,112 (99.0\%) \\
        Event end   & 1,706 & 0.592 & 1,577 (92.4\%) & 1,658 (97.2\%) & 1,637 (96.0\%) & 1,690 (99.1\%) \\
        \midrule
        Total       & 3,840 & 0.797 & 3,507 (91.3\%) & 3,732 (97.2\%) & 3,654 (95.2\%) & 3,802 (99.0\%) \\
        \bottomrule
    \end{tabular}
    }
    \vspace{5pt}
    \caption{Agreement statistics of event time labels across three annotators. ``MD'' denotes the mean pairwise absolute difference in frames between the labeled timestamps for each event. ``Exact3'' and ``Exact2'' indicate the number of events where at least three or two annotators, respectively, provided exactly the same timestamp. ``Close3'' and ``Close2'' indicate the number of events where at least three or two labels, respectively, fall within a two-frame (\SI{0.08}{s}) window.}
    \label{tab:annot}
\end{table}

\subsection{Baseline Comparison} \label{se:baseline}
Using the ground truth timestamps, we compared the performance of ELASTIC with a SOTA baseline, ETSY~\cite{VanRoyCD23}. As shown in Table~\ref{tab:accuracy}, ELASTIC significantly outperforms ETSY across all event categories, achieving 88.4\% exact alignment and 94.8\% within-second (W25) accuracy compared to ETSY's 21.1\% (Exact) and 74.7\% (W25), respectively. Additionally, unlike ETSY, ELASTIC accurately detects end frames of pass-like and set-piece events.

\begin{table}[t!]
    \centering
    \resizebox{\textwidth}{!}{
    \begin{tabular}{cc|rr|rrrrr}
        \toprule
        \textbf{Category} & \textbf{Method} & \makecell[c]{\bf Total} & \makecell[c]{\bf MD} & \makecell[c]{\bf Exact} & \makecell[c]{\bf W5} & \makecell[c]{\bf W25} & \makecell[c]{\bf W50} & \makecell[c]{\bf Valid} \\
        \specialrule{1pt}{2pt}{3pt}
        Pass-like & ETSY & 1,590 & 14.053 & 348 (21.9\%) & 995 (62.6\%) & 1,242 (78.1\%) & 1,375 (86.5\%) & 1,500 (94.3\%) \\
        & ELASTIC & 1,590 & \bftab{0.567} & \bftab{1,519} (\bftab{95.5\%}) & \bftab{1,546} (\bftab{97.2\%}) & \bftab{1,561} (\bftab{98.1\%}) & \bftab{1,566} (\bftab{98.5\%}) & \bftab{1,571} (\bftab{98.8\%}) \\
        \midrule
        Set-piece & ETSY & 117 & 11.974 & 68 (58.1\%) & 103 (88.0\%) & 107 (91.5\%) & 109 (93.2\%) & \bftab{116} (\bftab{99.2\%}) \\
        & ELASTIC & 117 & \bftab{0.123} & \bftab{108} (\bftab{92.3\%}) & \bftab{113} (\bftab{96.6\%}) & \bftab{114} (\bftab{97.4\%}) & \bftab{114} (\bftab{97.4\%}) & 114 (97.4\%) \\
        \midrule
        Incoming & ETSY & 168 & 24.181 & 9\phantom{..}(5.4\%) & 64 (38.1\%) & 107 (63.7\%) & 133 (79.2\%) & 160 (95.2\%) \\
        & ELASTIC & 168 & \bftab{6.584} & \bftab{144} (\bftab{85.7\%}) & \bftab{147} (\bftab{87.5\%}) & \bftab{151} (\bftab{89.9\%}) & \bftab{157} (\bftab{93.5\%}) & \bftab{166} (\bftab{98.8\%}) \\
        \midrule
        Minor & ETSY & 259 & 18.382 & 26 (10.0\%) & 82 (31.7\%) & 139 (53.7\%) & 162 (62.6\%) & 178 (68.7\%) \\
        & ELASTIC & 259 & \bftab{11.750} & \bftab{115} (\bftab{44.4\%}) & \bftab{174} (\bftab{67.2\%}) & \bftab{201} (\bftab{77.6\%}) & \bftab{233} (\bftab{90.0\%}) & \bftab{{248}} (\bftab{95.8\%}) \\
        \specialrule{1pt}{2pt}{3pt}
        Event start & ETSY & 2,134 & 15.154 & 451 (21.1\%) & 1,244 (58.3\%) & 1,595 (74.7\%) & 1,779 (83.4\%) & 1,954 (91.6\%) \\
        & ELASTIC & 2,134 & \bftab{2.346} & \bftab{1,886} (\bftab{88.4\%}) & \bftab{1,980} (\bftab{92.8\%}) & \bftab{2,027} (\bftab{95.0\%}) & \bftab{2,070} (\bftab{97.0\%}) & \bftab{2,099} (\bftab{98.4\%}) \\
        \midrule
        Event end & ELASTIC & 1,706 & 1.606 & 1,579 (92.6\%) & 1,623 (95.1\%) & 1,645 (96.4\%) & 1,661 (97.4\%) & 1,680 (98.5\%) \\
        \specialrule{1pt}{2pt}{3pt}
        Total & ELASTIC & 3,840 & 1.984 & 3,465 (90.2\%) & 3,603 (93.8\%) & 3,672 (95.6\%) & 3,731 (97.1\%) & 3,779 (98.4\%) \\
        \bottomrule
    \end{tabular}
    }
    \vspace{5pt}
    \caption{Synchronization accuracy of ETSY~\cite{VanRoyCD23} and ELASTIC (ours) across different event categories. ``Exact'' indicates the number of events exactly aligned to the ground truth, and ``W5/W25/W50'' count events whose predicted frame falls within 5/25/50 frames of ground truth. ``Valid'' refers to the number of events where the synchronizer successfully returned a single best frame.}
    \label{tab:accuracy}
\end{table}

Specifically, ELASTIC synchronizes pass-like events, which are both the most frequent (1,590 out of 2,134 events) and arguably the most important for downstream analysis, with near-perfect accuracy, achieving 95.6\% exact alignment and a mean difference of just 0.569 frames (\SI{0.023}{s}). While there remains some inaccuracy in minor event synchronization due to its inherent difficulty, ELASTIC still shows a substantial improvement over ETSY. Moreover, by isolating the detection of minor events into a separate stage, ELASTIC prevents their potential errors from propagating to other events. We analyze the effect of this design choice further in the ablation study presented in Section~\ref{se:ablation}. Overall, the precise and robust synchronization achieved by ELASTIC makes it a reliable foundation for various downstream tasks where accurate event timing is essential.

\subsection{Ablation Study} \label{se:ablation}
To investigate the effectiveness of our multi-stage strategy, we conducted an ablation study that compares the performance of different ordering of synchronization. Table~\ref{tab:ablation} presents results under three configurations: (1) synchronizing all events before receive detection, (2) delaying only minor event synchronization, and (3) delaying both incoming and minor events. Among these, configuration (2) that post-synchronizes only minor events after receive detection yields the best performance across both event start and end (receive) detection. This supports our design choice to isolate minor event synchronization, which reduces the risk of their inaccuracies propagating to the synchronization of other events.

In contrast, both configurations (1) and (3) exhibit a notable performance drop compared to our main configuration. The low accuracy of configuration (1) is because misaligned minor events can distort the temporal boundaries of the subsequent events, leading to cascading errors. Meanwhile, the degradation in configuration (3) is primarily due to the interdependence between receive detection and the timestamp of the following event. For instance, when a pass is followed by an incoming event, the accurate timestamp of that incoming event helps determine a precise window for detecting the reception of the pass. If the incoming event is not yet synchronized, the search window for the receive becomes overly broad, reducing the detection accuracy. The misaligned receive timing, in turn, negatively affects the synchronization of the subsequent incoming event.

Taken together, this findings suggest that incoming events, being both easier to detect due to their clear observable cues and tied to the reception of preceding events, should be detected before receive detection. In contrast, minor events, which are more error-prone and less relevant to previous receives, should be handled afterward in a separate final stage to maximize overall accuracy.

\begin{table}[t!]
    \centering
    \resizebox{\textwidth}{!}{
    \begin{tabular}{c|cc|rr|rrrrr}
        \toprule
        \textbf{Category} & \textbf{Incoming} & \textbf{Minor} & \makecell[c]{\bf Total} & \makecell[c]{\bf MD} & \makecell[c]{\bf Exact} & \makecell[c]{\bf W5} & \makecell[c]{\bf W25} & \makecell[c]{\bf W50} & \makecell[c]{\bf Valid} \\
        \midrule
        \multirow{3}{*}{\makecell{Event \\ start}}
        & & & 2,134 & 7.113 & 1,754 (82.2\%) & 1,858 (87.1\%) & 1,912 (89.6\%) & 1,971 (92.4\%) & 2,013 (94.3\%) \\
        & & \ding{51} & 2,134 & \bftab{2.346} & \bftab{1,886} (\bftab{88.4\%}) & \bftab{1,980} (\bftab{92.8\%}) & \bftab{2,027} (\bftab{95.0\%}) & \bftab{2,070} (\bftab{97.0\%}) & \bftab{2,099} (\bftab{98.4\%}) \\
        & \ding{51} & \ding{51} & 2,134 & 3.668 & 1,742 (81.6\%) & 1,857 (87.0\%) & 1,914 (89.7\%) & 1,972 (92.4\%) & 2,022 (94.8\%) \\
        \midrule
        \multirow{3}{*}{\makecell{Event \\ end}}
        & & & 1,706 & 1.679 & 1,488 (87.2\%) & 1,543 (90.5\%) & 1,565 (91.7\%) & 1,580 (92.6\%) & 1,599 (93.7\%) \\
        & & \ding{51} & 1,706 & \bftab{1.606} & \bftab{1,579} (\bftab{92.6\%}) & \bftab{1,623} (\bftab{95.1\%}) & \bftab{1,645} (\bftab{96.4\%}) & \bftab{1,661} (\bftab{97.4\%}) & 1,680 (98.5\%) \\
        & \ding{51} & \ding{51} & 1,706 & 10.079 & 1,483 (86.9\%) & 1,542 (90.4\%) & 1,576 (92.4\%) & 1,607 (94.2\%) & \bftab{1,683} (\bftab{98.7\%}) \\
        \bottomrule
    \end{tabular}
    }
    \vspace{5pt}
    \caption{Synchronization accuracy under different ordering configurations. A checkmark in the ``Incoming'' or ``Minor'' column indicates that the corresponding event category was synchronized after the receive detection.}
    \label{tab:ablation}
\end{table}

\section{Conclusions and Future Work} \label{se:conclusion}
This paper proposes ELASTIC, a framework for synchronizing event and tracking data in soccer without relying on noisy annotated event locations. By leveraging only tracking-derived motion features, ELASTIC is shown to improve the state-of-the-art synchronization accuracy by a large margin, including the detection of end times and receivers for pass-like events.

For future work, we plan to further refine the framework with extensive experiments. This includes hyperparameter tuning (e.g., scaling coefficients and clipping limits of scoring functions) and ablation studies on feature usage. Moreover, to assess the practical value of accurate synchronization, we will compare the performance of key soccer analytics tasks, including action selection prediction, pass success prediction, and expected possession value (EPV) estimation, on data before and after synchronization.

\section*{Acknowledgments}
This research was conducted using Eredivisie event and tracking data provided by AFC Ajax. We sincerely thank the club for their support.

%
%
%
\bibliographystyle{splncs04}
\bibliography{ref}

\end{document}